\documentclass[aps,amsmath,amssymb,twocolumn,superscriptaddress,nofootinbib]{revtex4-1}

\usepackage{graphicx}
\usepackage{dcolumn}
\usepackage{bm}
\usepackage{color}
\usepackage[colorlinks=true,pdfstartview=FitV,linkcolor=blue,citecolor=blue,urlcolor=blue]{hyperref}

\begin{document}

\newcommand{\sbfo}{\mbox{$^{124}$Sb}}
\newcommand{\ben}{\mbox{$^{9}$Be}}
\newcommand{\het}{\mbox{$^{3}$He}}
\newcommand{\fef}{\mbox{$^{55}$Fe}}
\newcommand{\eve}{\mbox{eV$_{\rm ee}$}}
\newcommand{\evr}{\mbox{eV$_{\rm nr}$}}
\newcommand{\drue}{\mbox{keV$_{\rm ee}^{-1}$}\mbox{\,kg}$^{-1}$\mbox{\,day}$^{-1}$}

\title{Measurement of the ionization produced by sub-keV silicon nuclear recoils\\ in a CCD dark matter detector}

\author{A.E.~Chavarria}
\email[]{alvaro@kicp.uchicago.edu}
\affiliation{Kavli Institute for Cosmological Physics and The Enrico Fermi Institute, The University of Chicago, Chicago, IL, United States}

\author{J.I.~Collar}
\affiliation{Kavli Institute for Cosmological Physics and The Enrico Fermi Institute, The University of Chicago, Chicago, IL, United States}

\author{J.R.~Pe\~{n}a}
\affiliation{Kavli Institute for Cosmological Physics and The Enrico Fermi Institute, The University of Chicago, Chicago, IL, United States}

\author{P.~Privitera}
\affiliation{Kavli Institute for Cosmological Physics and The Enrico Fermi Institute, The University of Chicago, Chicago, IL, United States}

\author{A.E.~Robinson}
\affiliation{Kavli Institute for Cosmological Physics and The Enrico Fermi Institute, The University of Chicago, Chicago, IL, United States}
\affiliation{Fermi National Accelerator Laboratory, Batavia, IL, United States}

\author{B.~Scholz}
\affiliation{Kavli Institute for Cosmological Physics and The Enrico Fermi Institute, The University of Chicago, Chicago, IL, United States}

\author{C.~Sengul}
\affiliation{Kavli Institute for Cosmological Physics and The Enrico Fermi Institute, The University of Chicago, Chicago, IL, United States}

\author{J.~Zhou}
\affiliation{Kavli Institute for Cosmological Physics and The Enrico Fermi Institute, The University of Chicago, Chicago, IL, United States}

\author{J.~Estrada}
\affiliation{Fermi National Accelerator Laboratory, Batavia, IL, United States}

\author{F.~Izraelevitch}
\affiliation{Fermi National Accelerator Laboratory, Batavia, IL, United States}

\author{J.~Tiffenberg}
\affiliation{Fermi National Accelerator Laboratory, Batavia, IL, United States}

\author{J.R.T.~de~Mello~Neto}
\affiliation{Universidade Federal do Rio de Janeiro, Instituto de  F\'{\i}sica, Rio de Janeiro, RJ, Brazil}

\author{D.~Torres~Machado}
\affiliation{Universidade Federal do Rio de Janeiro, Instituto de  F\'{\i}sica, Rio de Janeiro, RJ, Brazil}

\date{\today}

\begin{abstract}
We report a measurement of the ionization efficiency of silicon nuclei recoiling with sub-keV kinetic energy in the bulk silicon of a charge-coupled device (CCD). Nuclear recoils are produced by low-energy neutrons ($<$24\,keV) from a \sbfo-\ben\ photoneutron source, and their ionization signal is measured down to 60\,eV electron equivalent. This energy range, previously unexplored, is relevant for the detection of low-mass dark matter particles. The measured efficiency is found to deviate from the extrapolation to low energies of the Lindhard model. This measurement also demonstrates the sensitivity to nuclear recoils of CCDs employed by DAMIC, a dark matter direct detection experiment located in the SNOLAB underground laboratory.
\end{abstract}


\maketitle

\section{\label{sec:intro}Introduction}
Solid-state silicon detectors have been proposed for next-generation direct searches for dark matter~\cite{Aguilar-Arevalo:2016ndq, Rau:2012eg}. Thanks to their very low noise and the relatively low mass of the silicon nucleus, these detectors are most sensitive to low-mass ($<$10\,GeV/$\rm{c}^2$) weakly interacting massive particles (WIMPs), leading candidates to explain the nature of the Universe's dark matter~\cite{Kolb:1990vq, *Griest:2000kj, *Zurek:2013wia}. The dominant mechanism of interaction between WIMPs and baryonic matter is expected to be scattering off nuclei. Hence, WIMPs may be found from the ionization signal produced in the detector by the recoiling nucleus following a WIMP-nucleus interaction. 

At high recoil energies the ionization produced by a Si recoil approaches that of an electron of the same energy, while at very low Si recoil energies most of the energy goes into atomic collisions. Thus, the knowledge of the nuclear recoil ionization efficiency \textemdash\ the relation $E_e$$=$$\Gamma(E_r)$ between the energy deposited by the recoiling nucleus in the form of ionization, $E_e$, and  the nucleus kinetic energy, $E_r$ \textemdash\ is essential to establish the sensitivity of a WIMP detector and to characterize a potential signal.\footnote{Following convention, we express $E_r$ in units of \evr\ (eV nuclear recoil; kinetic energy deposited by a recoiling nucleus) and $E_e$ in units of \eve\ (eV electron equivalent; an electron of 5\,keV kinetic energy produces a 5\,k\eve\ signal through ionization).}
A low mass WIMP with typical speed of hundreds of km\,s$^{-1}$ will result in nuclear recoil energies $<$10\,k\evr. Theoretical models that predict the nuclear recoil ionization efficiency in Si are highly uncertain in this low-energy regime~\cite{Lindhard:1963vo, *ziegler1985stopping}, and dedicated calibrations are required.
Such measurements have been carried out in the past down to 3\,k\evr~\cite{PhysRevA.45.2104, *PhysRevD.42.3211, *PhysRevA.41.4058}. The next-generation of Si detectors for dark matter will achieve a detection threshold of $<$1\,k\evr\ where no experimental results are available.  
 
In this paper, we report a measurement of the nuclear recoil ionization efficiency in the range 0.7--2.3\,k\evr\ with neutrons from a \sbfo-\ben\ photoneutron source in a low-threshold silicon charge-coupled device (CCD).
Photoneutron sources have been used to characterize the response to nuclear recoils of NaI(Tl) scintillators~\cite{Collar:2013xva}, superheated fluids~\cite{alanrobinson} and P-type point contact germanium detectors~\cite{Scholz:2016qos}.
The CCD employed is a high-resistivity fully depleted device~\cite{1185186} developed in the R\&D efforts of DAMIC~\cite{Aguilar-Arevalo:2015lvd, Aguilar-Arevalo:2016ndq}, a dark matter direct detection experiment located in the  SNOLAB underground laboratory.

\section{\label{sec:setup}Experimental setup}

A \sbfo-\ben\ photoneutron source was used to produce low-energy neutrons: photonuclear interactions between 1691\,keV $\gamma$ rays from \sbfo\ decay and \ben\ result in almost monochromatic neutrons of 24\,keV.
The source consisted of a small rod of natural antimony, neutron activated at a nuclear reactor and then placed in a hollow BeO cylinder (American Beryllia, Inc.).
The initial activity of \sbfo\ was compared to a National Institute of Standards and Technology-traceable reference \sbfo\ $\gamma$ source using a high purity germanium detector, and found to be $4.76\pm0.20$~mCi.
The corresponding initial neutron production rate was estimated to be $\sim$10$^4$\,s$^{-1}$ from the known \ben$(\gamma,n)$ cross section and the geometry of the BeO target.
The neutron production rate decreased exponentially during the measurement period following the \sbfo\ decay rate ($\rm{T_{1/2}}=$ 60.20$\pm$0.01\,d~\cite{FLEMING1966251}).

 \begin{figure*}[t!]
\includegraphics[width=0.8\textwidth]{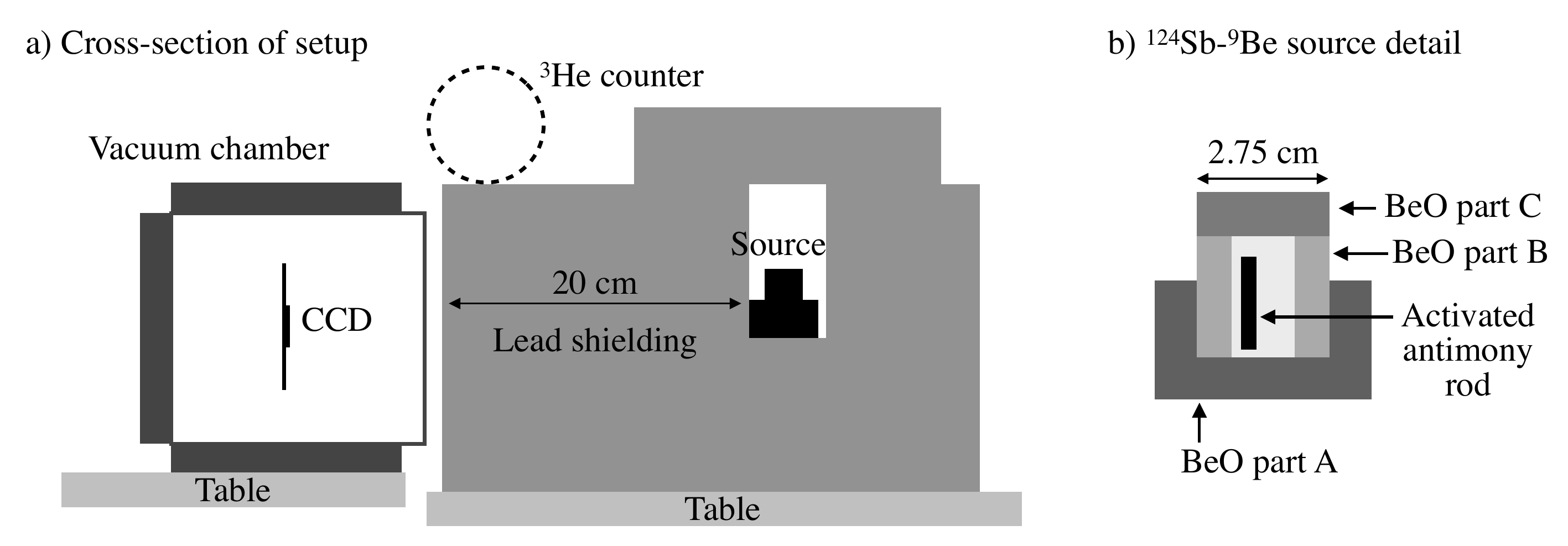}
\caption{\label{fig:setup} a)~Cross section of the experimental setup at the University of Chicago. A sample position of the \het\ counter used to validate the simulation of the neutron flux outside the lead shield is shown. b)~Cross section of the photoneutron source assembly. Materials other than BeO were also used to characterize the $\gamma$-ray background and validate the simulation of the neutron production in the source (see text).}
\end{figure*}
 
Neutrons from the source elastically scatter off silicon nuclei in the detector. The subsequent nuclear recoils deposit their kinetic energy ($<$3.2\,k\evr ) in the silicon bulk within 10\,nm of the interaction site, producing signals that mimic those expected from WIMP interactions.

The detector is an 8\,Mpixel CCD (pixel size 15$\times$15\,$\mu$m$^2$) with an active area of 18.8\,cm$^2$, a thickness of 500\,$\mu$m and a mass of 2.2\,g. The CCD was installed in a stainless-steel vacuum chamber ($10^{-6}$\,mbar) and cooled to the nominal operating temperature of 130\,K. The voltage biases, clocks and video signals required for the CCD operation were serviced by a Kapton flex cable wire bonded to the CCD. A bias of 128$\pm$1\,V was applied across the silicon substrate to fully deplete the substrate and minimize diffusion of charge carriers. There are no regions of partial or incomplete charge collection that may hinder the energy response of the device~\cite{Aguilar-Arevalo:2016ndq, 1185186}. The CCD was controlled and read out by commercial CCD electronics (Astronomical Research Cameras, Inc.). The pixel noise achieved with this system was 1.80$\pm$0.07\,$e^-$, equivalent to 6.8$\pm$0.3\,\eve\  (on average, 3.8\,\eve\ are required to produce a free charge carrier in silicon~\cite{4326950}).

The CCD is also sensitive to ionizing electrons produced by Compton scattering of $\gamma$ rays from the \sbfo\ source. The intense flux of $\gamma$ rays was suppressed to a level similar to that of environmental radiation by placing 20\,cm of lead between the source and the detector. Due to the large mass of the Pb nucleus and its small neutron capture cross section, the neutron flux was minimally attenuated by the lead shield, which acted mostly as a neutron radiator.

A dedicated measurement of the neutron flux was performed with an auxiliary \het\ counter (model LND-252) deployed in several positions around the lead shield and the CCD vacuum chamber. The \het\ counter was surrounded by 2.5\,cm of polyethylene to moderate fast neutrons, and by a 0.6-mm-thick cadmium layer to stop thermal neutrons. The typical uncertainty on the measurement of the neutron flux with this  \het\ counter was $\sim$5\%.  A diagram of the experimental setup, including a sample position of the \het\ counter and details of the photoneutron source, is shown in Fig.~\ref{fig:setup}.

\section{\label{sec:data}Data sets}

A total of 13268 images were acquired for this measurement between January and July, 2015. Data quality was excellent, with more than 99.5\% of the images deemed appropriate for analysis. The image exposure time was 972.17$\pm$0.08\,s. A large portion of the data was taken with the full BeO target (parts A, B and C in Fig.~\ref{fig:setup} made of 92, 30 and 14\,g of BeO, respectively), which provided the highest neutron rate.  Background $\gamma$-ray data were taken by replacing all the BeO target parts with geometrically identical pieces made of aluminum (Al), which has a similar $\gamma$-ray attenuation as BeO but a much higher photoproduction threshold, effectively turning off the neutron flux. Another data set with statistics equivalent to the full BeO target was taken with part B made of alumina ($\rm{Al_2 O_3}$) and the other parts of BeO.

Additional data were acquired with part A made of Al and part B made of BeO, and vice versa.  A number of images were taken with part A removed and a polyethylene capsule placed around the source to moderate the neutron spectrum. Data were also acquired with the source moved $\pm$3.4\,cm laterally within its lead shield. These additional data sets were used for crosschecks and systematic studies. 

\section{\label{sec:image}Image processing and Event reconstruction}

CCD data are two-dimensional images, where the pixel analog-to-digital converter (ADC) value is proportional to the number of charge carriers collected in the pixel. The charge from a single, pointlike ionization event may be spread over multiple adjacent pixels, due to the thermal diffusion of the carriers as they drift across the silicon substrate. The corresponding distribution over the pixel array is a two-dimensional (2D) Gaussian, with variance ($\sigma_{xy}^2$) proportional to the carrier transit time, and hence positively correlated with the depth of the interaction~\cite{Aguilar-Arevalo:2016ndq,1352164,*CeaseDiff}.

The image processing started with the determination of its pedestal value, corresponding to the dc offset introduced at the time of readout. 
The pedestal was estimated independently for each column of the image by a Gaussian fit of the ADC value distribution of the column's pixels, and then subtracted from every pixel value in the column. After this first column-based equalization, the same procedure was applied for each row of the image, yielding a final image in which the distribution of the pixel values is centered at zero with standard deviation equal to the pixel noise ($\sigma_{\rm{pix}}$). The presence of ionization events, which populate $<$1\% of pixels, does not introduce a bias in this procedure. Hot pixels or defects were identified as recurrent patterns over samples of hundreds of images and eliminated (``masked'') from the analysis ($<$5\% of the pixels were removed by this procedure). Since neutron-induced nuclear recoils result in small energy deposits, clusters of pixels with total energy $>$10\,k\eve\ were also masked (clusters were found as contiguous pixels with ionization signal greater than 4\,$\sigma_{\rm{pix}}$). The energy scale of the ionization signal was calibrated at 5.9\,k\eve\ using Mn K$_\alpha$ x rays from an \fef\ source, and confirmed to be linear down to 0.52\,k\eve\ using O fluorescence x rays. The linear response of the CCD has been demonstrated for signals as small as 10\,$e^-$~\cite{Aguilar-Arevalo:2016ndq}.

Low-energy deposits were searched throughout the image in a moving window of 7$\times$7 pixels. The size of the window was chosen to contain the charge spread expected for pointlike energy deposits. For a given window's position, the difference in log-likelihood ($\Delta LL$) between two hypotheses \textemdash\ the first of a 2D Gaussian distribution of charge on top of white noise, the second of only white noise \textemdash\ was calculated. If the 2D Gaussian hypothesis was found more likely, the window was moved around to find the local $\Delta LL$ maximum, to properly center the event in the window. Then a fit was performed from which the $x$-$y$ position, charge spread and energy ($E$) of the candidate ionization event were obtained as the best-fit values of the center ($\vec{\mu}$), standard deviation ($\sigma_{xy}$) and integral of the 2D Gaussian, respectively. No further selection criteria were applied to events with reconstructed $E$$>$0.25\,k\eve . Lower-energy candidates were required to have $\Delta LL$$<$$-22.5$ to guarantee $<$0.01 accidental events from noise per image. Also, the distance between $\vec{\mu}$ and the center of the fitting window was requested to be $<$1.5\,pixels, to avoid clipped events that may occur when multiple ionization events are too close together.

The event selection procedure was validated by introducing simulated events on top of a representative subsample of 1000 raw data images. The simulated events were parametrized as 2D Gaussian distributions of pixel values with amplitudes and spreads as expected for pointlike ionization events distributed uniformly in the silicon bulk. The event selection was then applied to this set of images. The window fitting procedure was found to properly extract the parameters of the simulated events. In particular, for events in the energy interval relevant for nuclear recoils, the simulated ionization signal was reconstructed with a resolution equivalent to 20\,\eve .
Thus, the energy response was modeled with a resolution $\sigma_E^2$$=$$(20{\rm \,eV_{ee}})^2$$+$$(3.77{\rm \,eV_{ee}})FE$, where $F$ is the Fano factor~\cite{PhysRev.72.26, *doi:10.1117/12.948704}.
As the energy response could not be well characterized for events with $E$$<$60\,\eve\ due to the small significance of the signal over image noise, these events were excluded from the analysis. This set of simulations was also used to estimate the event selection acceptance as a function of energy, which was found to be around (20$\pm$1)\% at 60\,\eve\ and reaching 100\% at 150\,\eve. The smaller acceptance at lower energies is due to a loss in sensitivity to interactions that occur deeper in the bulk and closer to the back of the device, for which lateral charge diffusion is significant and the collected charge is spread over a larger number of pixels, leading to a smaller significance of the pixel signal over noise. The diffusion model implemented in the simulation and the procedure to estimate the event acceptance have been validated with Compton-scattered electrons from $\gamma$-ray calibration data in Ref.~\cite{Aguilar-Arevalo:2016ndq}.

\section{\label{sec:spec}Energy spectrum and Simulations}

\begin{figure}[b!]
\includegraphics[width=0.49\textwidth]{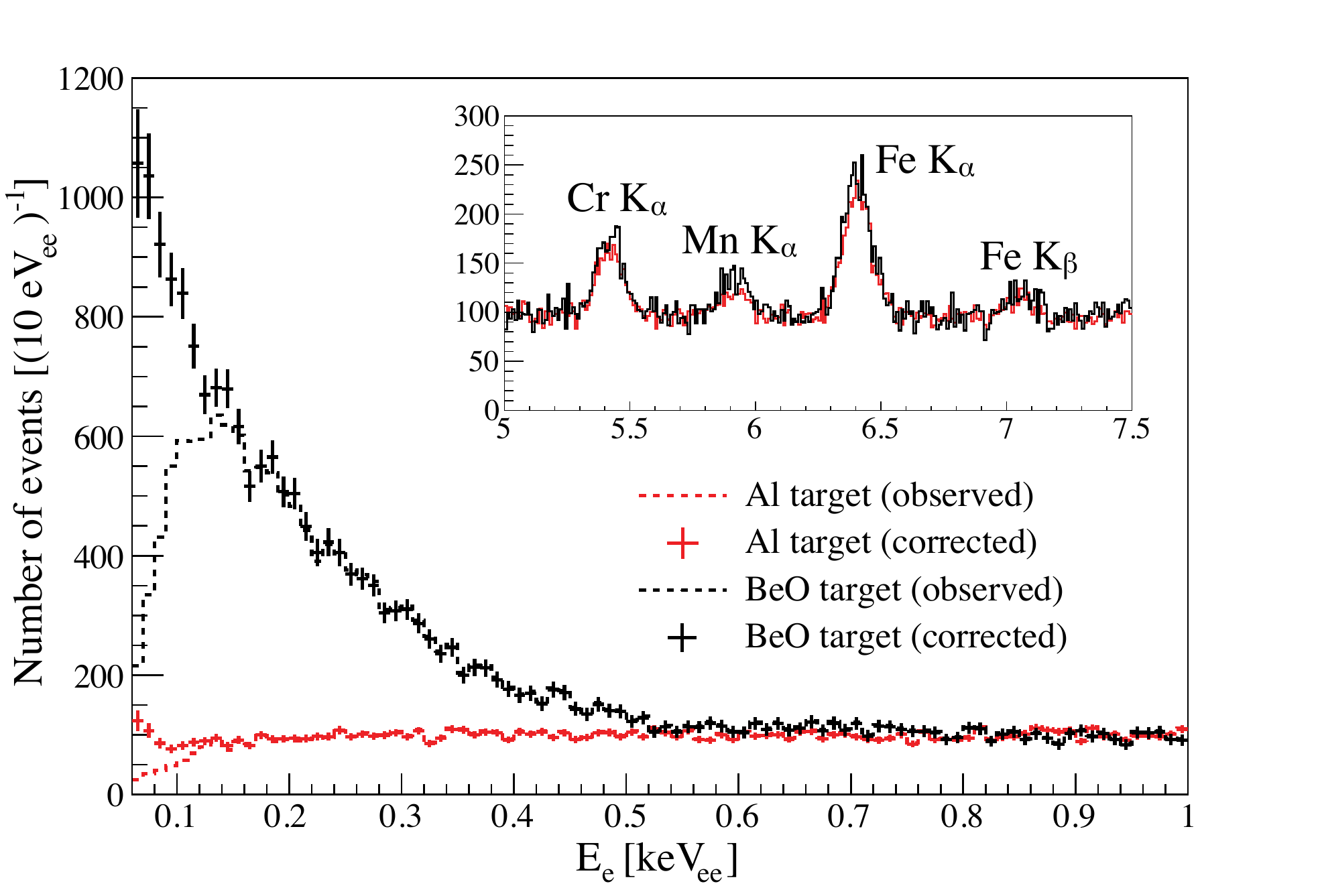}
\caption{\label{fig:sig_bkgd}Measured ionization spectra with the full BeO and Al targets (dashed lines). Solid markers represent the spectra corrected for the energy-dependent event selection acceptance. The inset shows the spectra in the 5.0--7.5\,k\eve\ range, with in-run calibration lines from fluorescence x rays originating in the stainless steel of the vacuum chamber.}
\end{figure}

The measured ionization spectra from the data sets with the full BeO (neutrons on) and Al (neutrons off) targets are shown in Fig.~\ref{fig:sig_bkgd}, with the Al target spectrum normalized to the time period of the BeO target data set. Since a component of the background is expected to be proportional to the \sbfo\ source intensity, the normalization was derived by measuring the rate of events with energy between 2 and 5\,k\eve\ as a function of time. This rate was well fitted by the sum of an exponential with decay time consistent with that of the \sbfo\ source, and a constant component associated with environmental background. The Al target spectrum was then normalized taking into account the intensity of the source during the full BeO target data set. A clear excess of events from neutron-induced nuclear recoils was observed below 0.5\,k\eve\ in the full BeO target spectrum. The nuclear recoil ionization spectrum was then obtained by subtracting the full Al target spectrum representative of the $\gamma$-ray background. Note that since the spatial resolution of the CCD ($\sim$$\mu$m) is much smaller than the neutron interaction length in silicon, each ionization event is the result of a single recoil, which simplifies the interpretation of the spectrum. 
Calibration lines from fluorescence of Fe, Cr and Mn in the stainless steel of the vacuum chamber (inset of Fig.~\ref{fig:sig_bkgd}) were used to monitor the energy scale, which was stable to $<$1\% over the six months of data taking.

\begin{figure}[b!]
\includegraphics[width=0.49\textwidth]{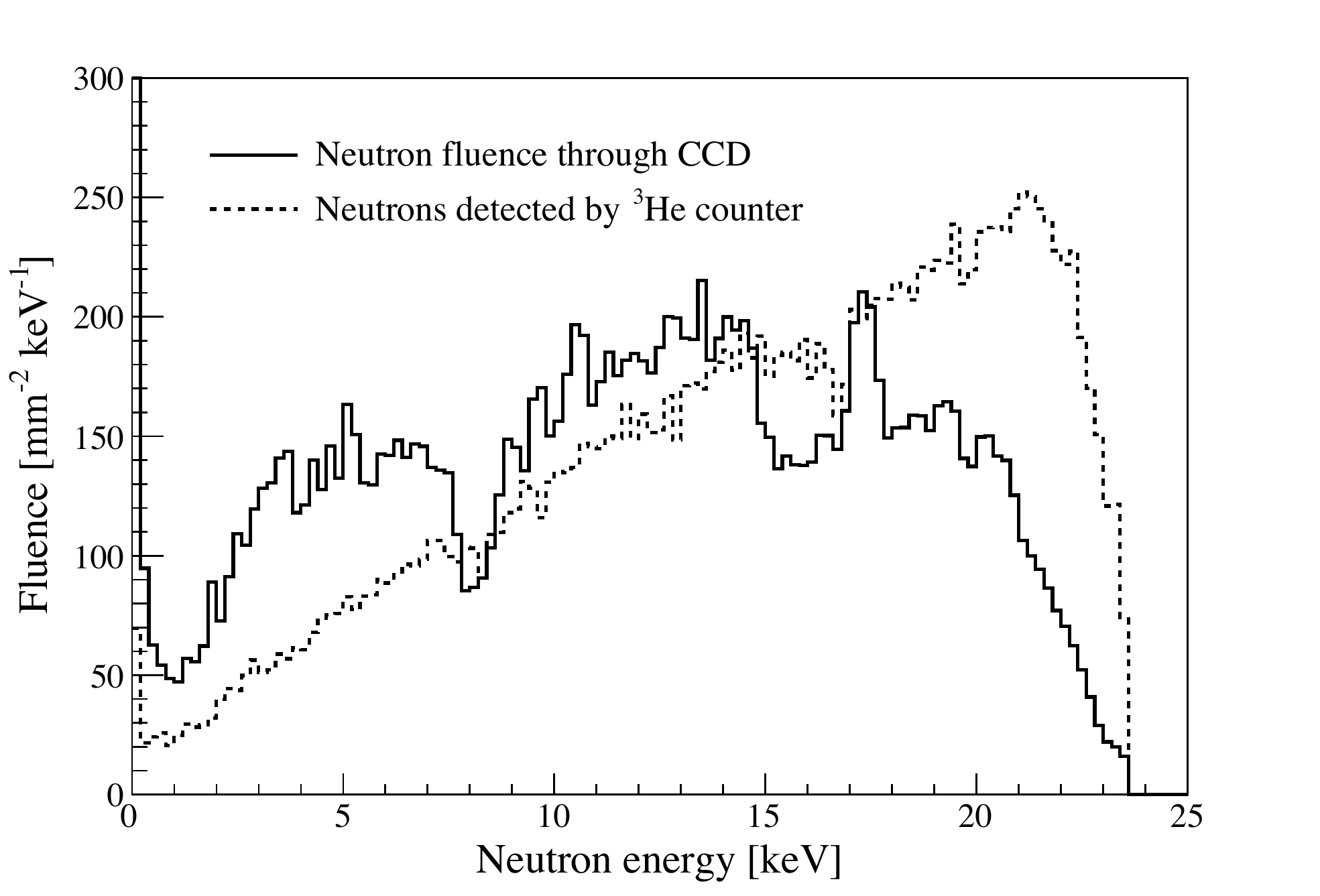}
\caption{\label{fig:fluence}Simulated neutron fluence through the CCD with the full BeO target (solid line). The monochromaticity of the original 24\,keV neutrons is lost due to moderation in the source materials and in the lead shield. The simulated energy spectrum of neutrons detected by the \het\ counter (dashed line) is also shown, with an arbitrary normalization for illustration.}
\end{figure}

A detailed MCNP5 simulation~\cite{mcnp} of the experiment was performed, including the full setup geometry and photoneutron production in the BeO target. The latest cross sections for neutron elastic scattering in Pb~\cite{PhysRevC.89.032801} were implemented in the MCNP libraries. The simulated neutron fluence through the CCD is shown in Fig.~\ref{fig:fluence}. Note that the initially monochromatic photoneutron energy is degraded by scattering in the BeO, in the lead shield and in materials surrounding the CCD; structures in the neutron spectrum are caused by material-dependent resonances in the neutron cross section. A nuclear recoil energy ($E_r$) spectrum was then calculated from the neutron fluence using the total and differential neutron-silicon scattering cross sections given in Ref.~\cite{larson76, *PhysRev.126.1105}. For a proper comparison with the data, the $E_r$ spectrum was normalized to the number $N_D$ of \sbfo\ decays calculated from the measured $\gamma$-ray activity and the exposure time of the data set. A small correction to take into account the number of masked pixels in each image was applied. 
Simulations corresponding to all the data sets detailed in Sec.~\ref{sec:setup} were performed. 

The \het\ counter's response was also simulated with MCNP5. Fig.~\ref{fig:fluence} shows the energy spectrum of neutrons detected by the counter in one of its locations. This spectrum covers reasonably well the energy range of the neutrons entering the CCD. Simulations were performed for all locations of the \het\ counter.

\section{\label{sec:res}Analysis methods}

\begin{figure}[b!]
\includegraphics[width=0.49\textwidth]{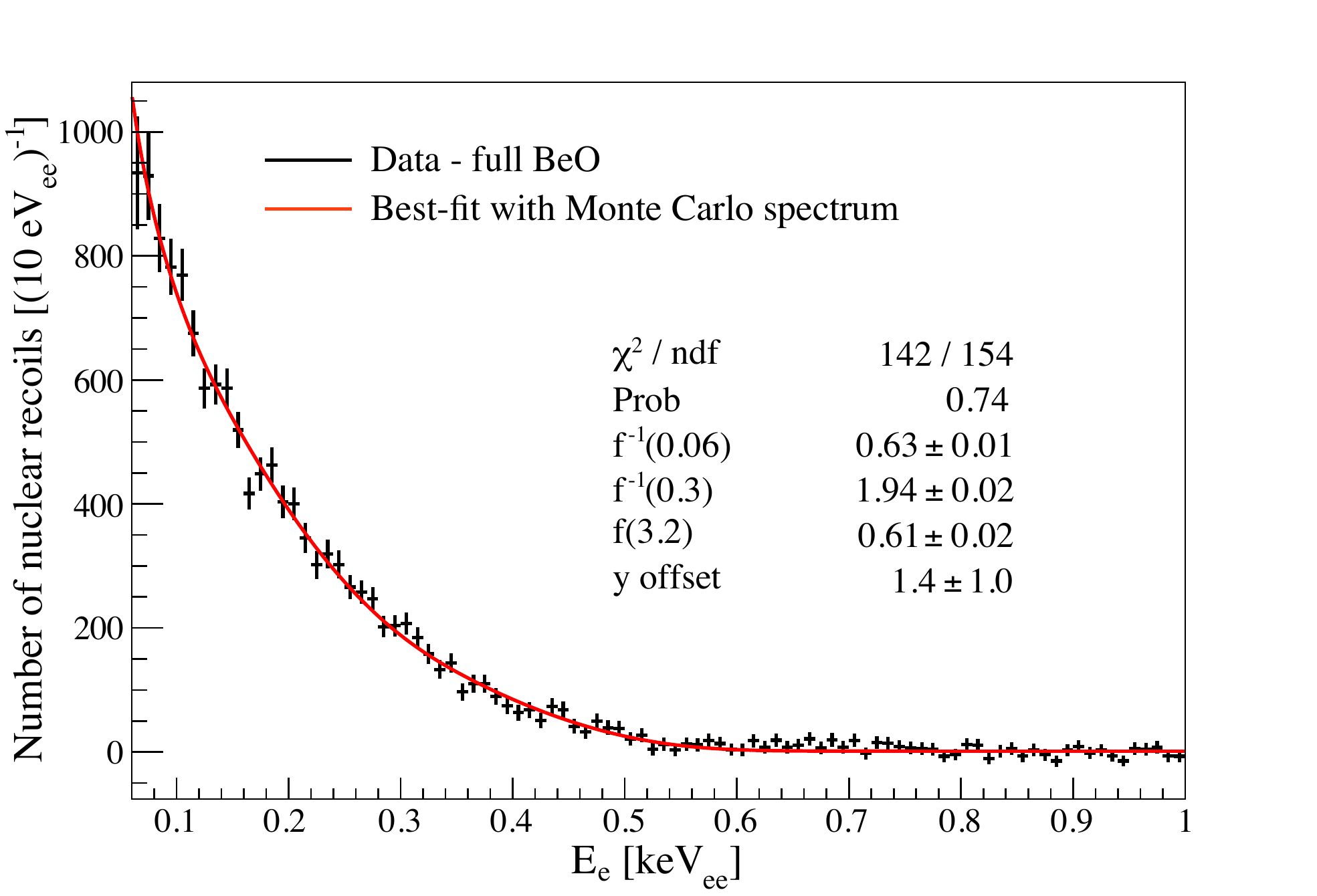}
\caption{\label{fig:mc_fit} Ionization spectrum of nuclear recoils induced by neutrons from the full BeO target source (black markers) and best fit to the data (solid line). The fitting function was obtained by applying a cubic spline model $f$ of the nuclear recoil ionization efficiency to the simulated recoil spectrum and convolving with the detector energy resolution. The best-fit parameters of the spline are given in the legend.}
\end{figure}

Two independent methods were used to determine the nuclear recoil ionization efficiency, i.e., the relation $E_e$$=$$\Gamma(E_r)$. 
In the first method, the energy deposited by the nucleus in the form of ionization $E_e$ was parametrized as a function of the nuclear recoil energy $E_r$ by a cubic spline function $f$ with three free parameters, i.e., $E_e/\mbox{k\eve}$$=$$f(E_r/\mbox{k\evr})$. 
 An $E_e$ spectrum to be fitted to the data was then derived by applying the spline function to the  simulated $E_r$ spectrum of Sec.~\ref{sec:spec} and convolving with the energy resolution. The best-fit spline from a $\chi^2$ minimization provided a first estimate of the nuclear recoil ionization efficiency. A constant offset in the spectrum baseline was also left free in the fit to account for a possible residual background, and found to be consistent with zero at 1.4\,$\sigma$. Fig.~\ref{fig:mc_fit} shows the result of the fit to the data set with the full BeO target.

\begin{figure}[b!]
\includegraphics[width=0.49\textwidth]{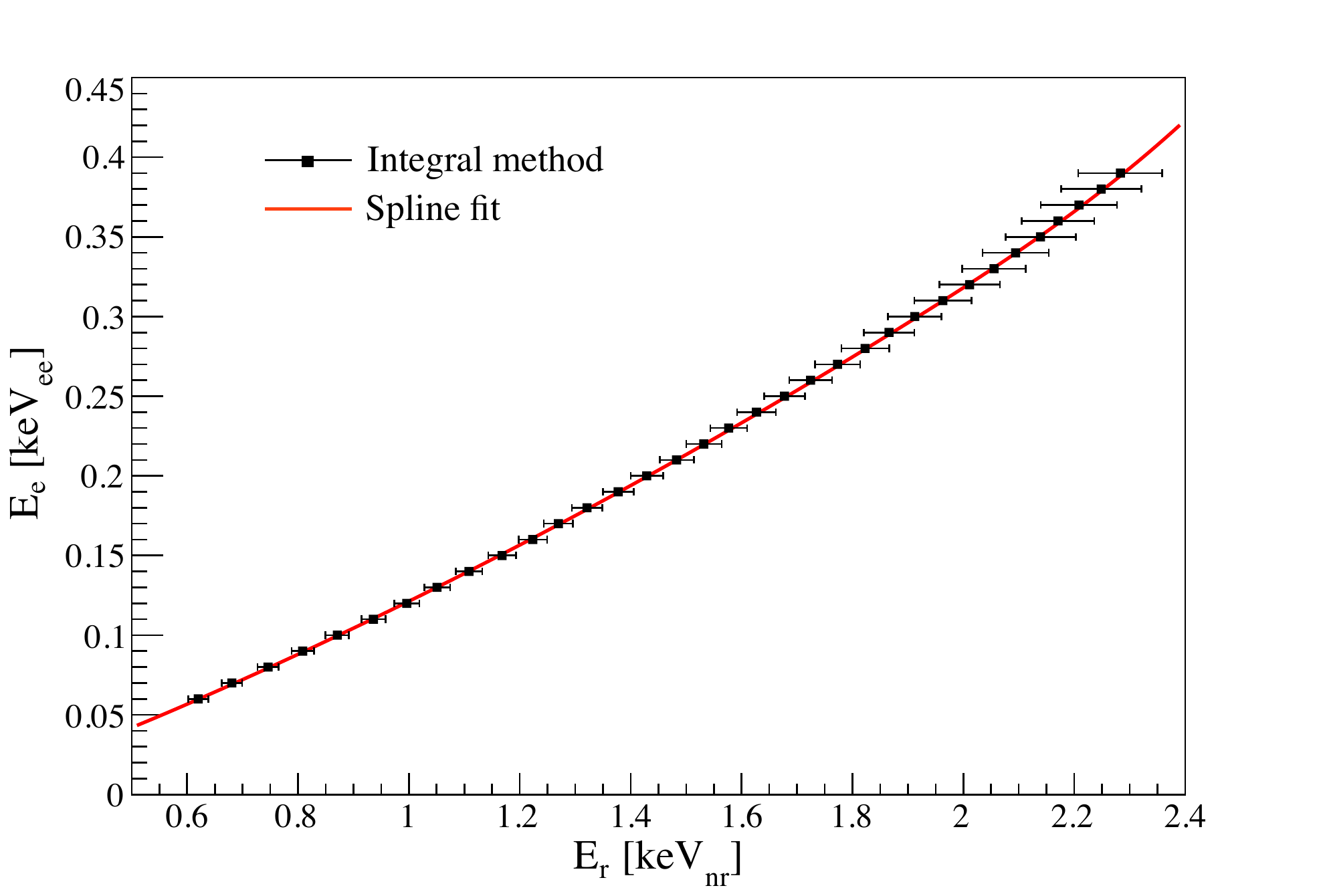}
\caption{\label{fig:compare} The nuclear recoil ionization efficiency obtained by applying the integral method (black markers) and the spline fit (solid line) to the full BeO target data. Statistical uncertainties obtained by the integral method are also shown.}
\end{figure}

A second method which does not assume a functional form was used for the final result and to evaluate statistical and systematic uncertainties. It is based on the fact that the integral number of events above a value $E^*_e$ in the measured nuclear recoil ionization spectrum must be equal to that of the simulated nuclear recoil energy spectrum above a recoil energy $E^*_r$ when $E^*_e$$=$$\Gamma(E^*_r)$. First, the number of events $N_e(E_e$$\ge$$E^*_e)$ with ionization signal above a given $E^*_e$ (defined as the lower edge of a spectrum bin) was calculated. Then, the corresponding recoil energy $E^*_r$ was found for which $N_r(E_r$$\ge$$E^*_r)$$=$$N_e(E_e$$\ge$$ E^*_e)$, $N_r(E_r$$\ge$$E^*_r)$ being the number of events with $E_r$$\ge$$E^*_r$ in the simulated nuclear recoil energy spectrum. The ensemble of pairs $(E^*_e, E^*_r)$ determined the nuclear recoil ionization efficiency. Before applying the integral method to the data, a possible residual background in the measured spectrum was taken into account. A background contribution which is only present with the BeO target \textemdash\ for example, $\gamma$ rays produced by neutron capture in the material surrounding the CCD \textemdash\ would still remain after the Al target spectrum subtraction. The small positive constant offset obtained in the spline fit (Fig.~\ref{fig:mc_fit}) may be an indication of such background. An estimate of the residual background was obtained from the events in the 1.0--1.6\,k\eve\ range where no nuclear recoil signal is expected, and the nuclear recoil ionization spectrum was corrected accordingly. 
In the simplified description given above, the integral method does not take into account that events may be spread over several bins by the resolution in the ionization signal. An iterative procedure was used to include this effect. In iteration $i$, an $E_e$ spectrum $S_i(E_e)$ was constructed by applying the nuclear recoil ionization efficiency $\Gamma_{i-1}$ from iteration $(i$$-$$1)$ to the simulated $E_r$ spectrum and convolving with the energy resolution.
The efficiency $\Gamma_{i-1}$ was used to convert the $S_i(E_e)$ spectrum back to a spectrum $S_i(E_r)$. An efficiency $\Gamma_i$ was then derived by applying the integral method to the data using the $S_i(E_r)$ spectrum. Less than five iterations were required for a convergence of the method.  The integral method was validated by applying it on mock data sets generated with known nuclear recoil ionization efficiencies.  
The nuclear recoil ionization efficiency obtained applying the integral method to the full BeO target data is compared in Fig.~\ref{fig:compare} with the spline fit showing excellent agreement and providing further confidence in the result. 

Statistical uncertainties on $E^*_r$ were evaluated by repeating the procedure for $N_e \pm \Delta N_e$, where $\Delta N_e$ accounts for the statistical uncertainty of the measured spectrum, including the $\gamma$-ray background subtraction.

\section{\label{sec:res}Results}

The integral method was applied to the data sets with different source configurations (Sec.~\ref{sec:setup}) and a weighted average of the corresponding nuclear recoil ionization efficiencies was taken as the best estimate (Table~\ref{tab:ie}). The result is determined by the two high statistics samples, the full BeO target and the alumina$+$BeO target, with other data sets consistent with at least one of these within statistical uncertainty.

Several sources of systematic uncertainty were evaluated. The nuclear recoil ionization efficiency of the full BeO target data is generally in good agreement with that of the alumina$+$BeO target \textemdash\ the maximum deviation occurring at $E_e$$=$60\,\eve\ where the corresponding recoil energies differ by 16\% \textemdash\ but differences at low energy are beyond statistical uncertainty. This may be due to an imperfect simulation of the production and moderation of neutrons in the source. The difference between the average and the nuclear recoil ionization efficiency of the two high statistics samples was taken as an estimate of the associated systematic uncertainty ranging from $<$1\% at $E_e$$=$390\,\eve\ to  8\% at $E_e$$=$60\,\eve .

\begin{table}[b!]
\caption{\label{tab:ie}Measured ionization signal ($E_e$) as a function of nuclear recoil energy ($E_r$). Statistical and systematic uncertainties are also quoted. As noted in the text, the uncertainties are strongly correlated between energy bins.}
\begin{ruledtabular}
\begin{tabular}{cccc}
$E_e$&
$E_r$&
Statistical&
Systematic\\
(keV$_{\textrm{ee}}$) &
(keV$_{\textrm{nr}}$) &
uncertainty (keV$_{\textrm{nr}}$) &
uncertainty (keV$_{\textrm{nr}}$) \\
\colrule
0.06 & 0.68 & $\pm$0.01 & $\pm$0.10\\
0.09 & 0.86 & $\pm$0.01 & $\pm$0.10\\
0.12 & 1.05 & $\pm$0.01 & $_{-0.09}^{+0.10}$\\
0.15 & 1.22 & $\pm$0.02 & $_{-0.08}^{+0.09}$\\
0.18 & 1.36 & $\pm$0.02 & $_{-0.07}^{+0.08}$\\
0.21 & 1.52 & $\pm$0.02 & $\pm$0.07\\
0.24 & 1.66 & $\pm$0.02 & $\pm$0.06\\
0.27 & 1.81 & $_{-0.03}^{+0.02}$ & $\pm$0.06\\
0.30 & 1.94 & $\pm$0.03 & $_{-0.05}^{+0.06}$\\
0.33 & 2.07 & $_{-0.04}^{+0.03}$ & $\pm$0.04\\
0.36 & 2.18 & $\pm$0.04 & $\pm$0.04\\
0.39 & 2.28 & $\pm$0.05 & $\pm$0.03\\
\end{tabular}
\end{ruledtabular}
\end{table}

Another important systematic uncertainty comes from the knowledge of the simulated recoil spectrum, in particular its absolute normalization. Directly relevant to the recoil spectrum normalization are the \ben$(\gamma,n)$ cross section and the neutron-silicon scattering cross section. 
We use $1.41$$\pm$$0.04$\,mb for the \ben$(\gamma,n)$ cross section at a photon energy of 1691\,keV obtained from a recent reanalysis of radioisotope measurements~\cite{PhysRevC.94.024613} that is consistent with the latest inverse Compton-scattering photon beam experiment~\cite{Utsunomiya}.
The neutron-silicon scattering cross section is known with an uncertainty of 3\%~\cite{larson76}.
The uncertainty in the number of \sbfo\ decays $N_D$ was dominated by the 4\% uncertainty in the intensity of the \sbfo\ $\gamma$ source, with the uncertainty in the exposure time being negligible.

The quality of the simulation was validated by the neutron rate measurements with the \het\ counter. The detector was placed in seven positions around the lead shield and the vacuum chamber, with counting rates varying by up to a factor of 25 when shielding from the source increased from 8 to 20\,cm of lead. The ratio between the measured and predicted neutron rates over the seven positions was 1.10$\pm$0.07. Since the statistical uncertainty of the rate measurements was $<$1\%, the observed rms of the ratio (0.07) was taken as an estimate of the systematic uncertainty associated with the simulation of the geometry of the apparatus and of the neutron propagation. Note that the absolute value of the ratio (1.10) is compatible with unity within its uncertainty [$\pm$0.08 estimated from the uncertainty on the \ben$(\gamma,n)$ cross section, the absolute intensity of the \sbfo\ source and the \het\ counter], providing an independent check of the normalization of the neutron fluence. 

By summing these uncertainties in quadrature, a total systematic uncertainty of 8.5\% was assigned to the magnitude of the simulated spectrum. The integral method analysis was then repeated scaling the normalization of the simulated spectrum by $\pm 8.5\%$, and the differences in the nuclear recoil ionization efficiency were taken as an estimate of the corresponding systematic uncertainty.

Other potential sources of systematic uncertainty were investigated.
The Fano noise~\cite{PhysRev.72.26, *doi:10.1117/12.948704} in the response of a silicon detector to ionization produced by low-energy nuclear recoils is unknown. We repeated the analysis including in the energy resolution a Fano factor ranging from 0.15 (as for fully ionizing particles) to unity, with negligible effect on the result.
The influence of background in the measured recoil ionization spectrum was evaluated by repeating the analysis without performing the residual background subtraction (Sec.~\ref{sec:res}), which resulted in changes of the nuclear recoil ionization efficiency $<$1\%.
Subdominant 2091\,keV $\gamma$ rays from \sbfo\ decay also produce neutrons. We performed a dedicated simulation of the recoil energy spectrum including this neutron component, which contributes $<$1\% to the nuclear recoil signal below 1\,k\eve, and found it to have a negligible effect on the extracted nuclear recoil ionization efficiency. In the integral method, the number of events above a given energy $E_e^*$ were calculated summing in the range $E_e^*$$\le$$E_e$$\le$1\,k\eve . Changing the upper limit of the integration to 0.7\,k\eve\ resulted in $<$1\% differences in the nuclear recoil ionization efficiency. 
 
\begin{figure}[t!]
\includegraphics[width=0.49\textwidth]{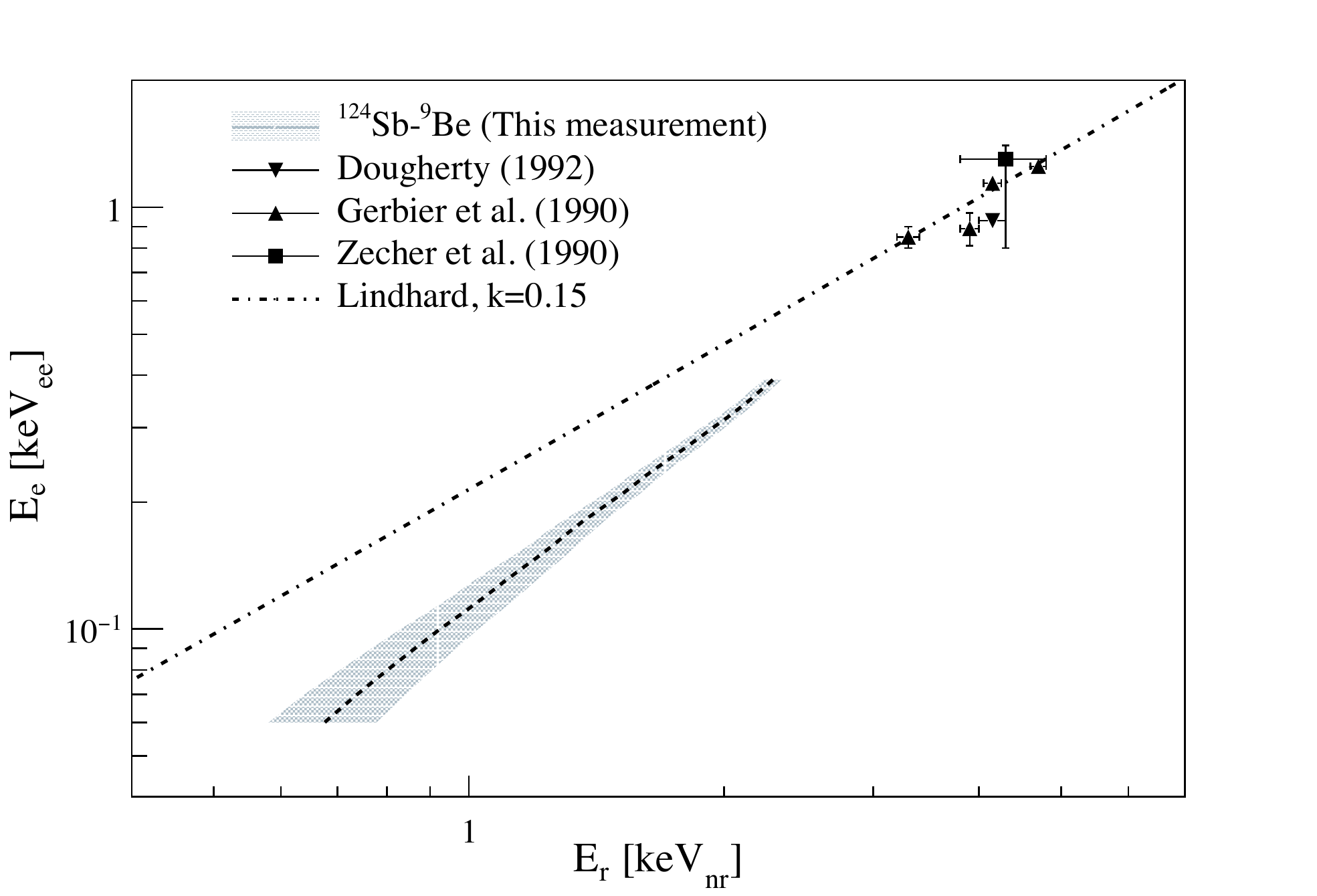}
\caption{\label{fig:ie_vs_er}
Ionization signal ($E_e$) as a function of nuclear recoil energy ($E_r$) in silicon. The gray band represents the 1\,$\sigma$ uncertainty in our measurement. Results from previous experiments and the extrapolation of the theoretical Lindhard model are shown for reference.}
\end{figure}

Results are summarized in Table~\ref{tab:ie}. Note that the quoted uncertainties are to a good approximation fully correlated by the nature of the integral method. The measured nuclear recoil ionization efficiency is shown in Fig.~\ref{fig:ie_vs_er}, where the uncertainty band is obtained from the quadrature sum of statistical and systematic uncertainties. Also shown are results from previous experiments~\cite{PhysRevA.45.2104} and the extrapolation of the theoretical Lindhard model with $k$$=$0.15, as predicted for silicon~\cite{Lindhard:1963vo, *ziegler1985stopping}.

\section{\label{sec:conc}Conclusions}

We reported a measurement of the nuclear recoil ionization efficiency in silicon for nuclear recoils between 0.7 and 2.3\,k\evr, a range previously unexplored and relevant for the direct detection of low-mass WIMPs. The measured efficiency was found to deviate significantly from the extrapolation to low energies of Lindhard model, which has been usually employed in sensitivity forecasts of next-generation experiments. Recent measurements with fast neutrons at higher recoil energies~\cite{antonella2, *antonella1} are also consistent with a deviation from Lindhard model in the range between our result and 3\,k\evr. 

\begin{figure}[t!]
\includegraphics[width=0.49\textwidth]{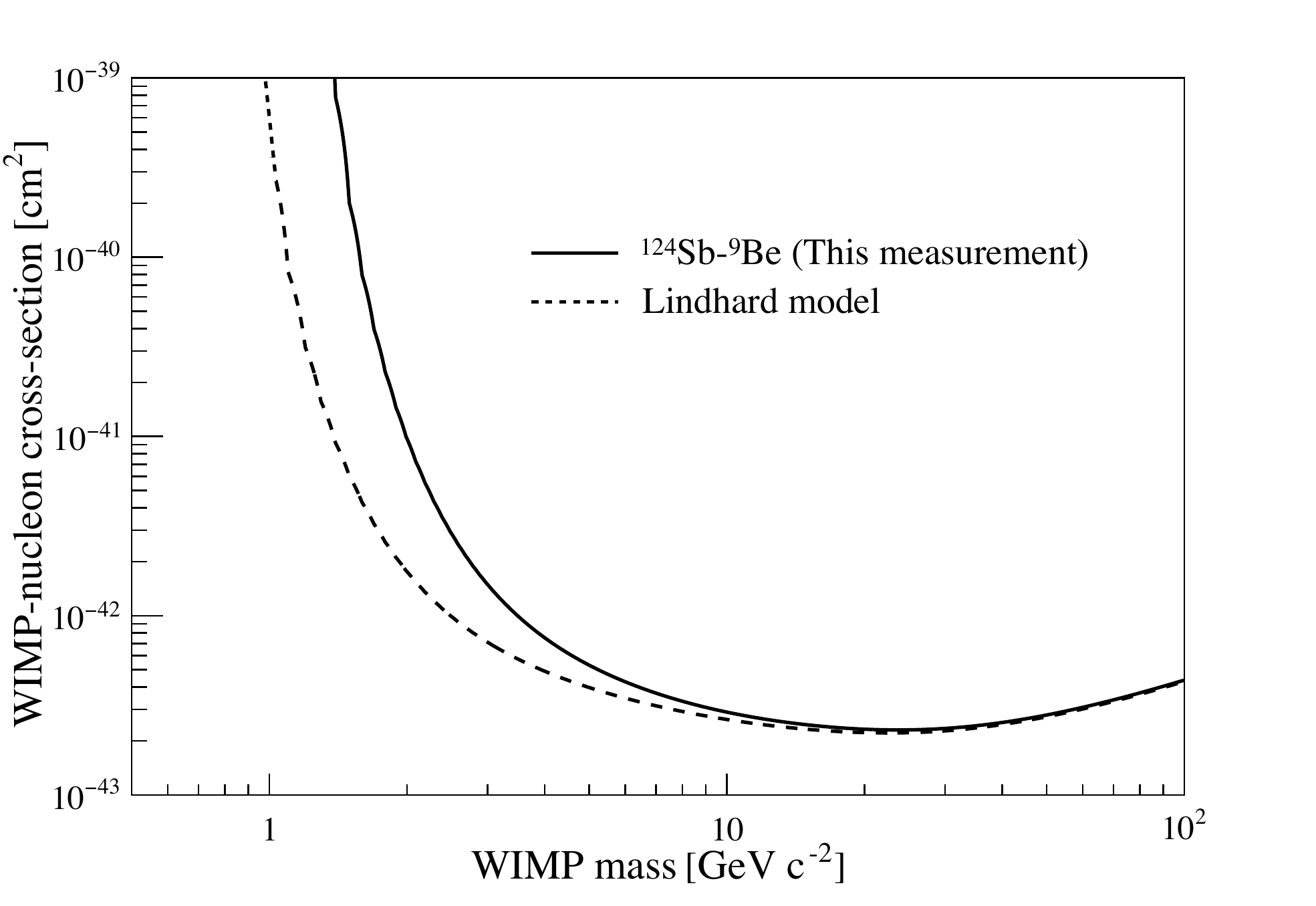}
\caption{\label{fig:forecast}Excluded region of the spin independent WIMP-nucleon elastic scattering cross section and WIMP mass (90\% C.L.) forecast for a generic silicon detector (see text) using the nuclear recoil ionization efficiency measured in this work (solid line) and the Lindhard model (dashed line).}
\end{figure}

To illustrate the impact of our measurement on upcoming experiments~\cite{Aguilar-Arevalo:2016ndq,Rau:2012eg} we show in Fig.~\ref{fig:forecast} the sensitivity of a generic silicon detector assuming a 100-kg-day background-free exposure with an energy threshold of 60\,\eve . When the Lindhard model is used, the sensitivity at the lowest WIMP masses is overstated. 

This measurement proved the capability of DAMIC CCDs to detect potential sub-k\evr\ signals from WIMPs. The detector was operated with an unprecedentedly low-energy threshold, high duty cycle and stable noise and energy response, collecting more than $10^4$ images over a period of six months \textemdash\ all important requirements for a successful dark matter search.
   
\begin{acknowledgments}
This work has been supported by the Kavli Institute for Cosmological Physics at the University of Chicago through Grant No. NSF PHY-1125897 and an endowment from the Kavli Foundation, and by Grant No. NSF PHY-1506208. We are grateful to the following agencies and organizations for financial support: Coordena\c{c}\~{a}o de Aperfei\c{c}oamento de Pessoal de N\'{\i}vel Superior (CAPES), Conselho Nacional de Desenvolvimento Cient\'{\i}fico e Tecnol\'{o}gico (CNPq) and  Funda\c{c}\~{a}o de Amparo \`{a} Pesquisa do Estado de Rio de Janeiro (FAPERJ), Brazil. We thank American Beryllia, Inc. for providing the beryllium oxide target.
\end{acknowledgments}

\bibliography{myrefs.bib}

\end{document}